\documentclass{article}

\usepackage[final]{nips_2017}

\usepackage{siunitx}

\usepackage[utf8]{inputenc} 
\usepackage[T1]{fontenc}    
\usepackage{hyperref}       
\usepackage{url}            
\usepackage{booktabs}       
\usepackage{amsfonts}       
\usepackage{nicefrac}       
\usepackage{microtype}      

\usepackage{graphicx}
\usepackage{caption}
\usepackage{subcaption}

\title{3D Deep Learning for Biological Function Prediction from Physical Fields}

\author{
Vladimir~Golkov\,$^{\text{1}}$,
Marcin~J.~Skwark\,$^{\text{2}}$,
Atanas~Mirchev\,$^{\text{1}}$,
Georgi~Dikov\,$^{\text{1}}$,\\
\textbf{
Alexander~R.~Geanes\,$^{\text{2}}$,
Jeffrey~Mendenhall\,$^{\text{2}}$,
Jens~Meiler\,$^{\text{2}}$ and
Daniel~Cremers\,$^{\text{1}}$
}
\\
$^1$ Technical University of Munich, Germany\\
$^2$ Vanderbilt University, Nashville, TN, USA\\
\footnotesize{\texttt{ golkov@cs.tum.edu, marcin@skwark.pl, \{atanas.mirchev, georgi.dikov\}@tum.de, }} \\
\footnotesize{\texttt{ 
\{alexander.r.geanes, jeffrey.l.mendenhall, jens.meiler\}@vanderbilt.edu, }} \\
\footnotesize{\texttt{ 
cremers@tum.de  }}
}

\begin{document}

\maketitle

\begin{abstract}
Predicting the biological function of molecules, be it proteins or drug-like compounds, from their atomic structure is an important and long-standing problem. Function is dictated by structure, since it is by spatial interactions that molecules interact with each other, both in terms of steric complementarity, as well as intermolecular forces. Thus, the electron density field and electrostatic potential field of a molecule contain the ``raw fingerprint'' of how this molecule can fit to binding partners. 
In this paper, we show that deep learning can predict biological function of molecules \emph{directly from their raw 3D approximated electron density and electrostatic potential fields}. Protein function based on EC numbers is predicted from the approximated electron density field. In another experiment, the activity of small molecules is predicted with quality comparable to state-of-the-art descriptor-based methods.
We propose several alternative computational models for the GPU with different memory and runtime requirements for different sizes of molecules and of databases.
We also propose application-specific multi-channel data representations. With future improvements of training datasets and neural network settings in combination with complementary information sources (sequence, genomic context, expression level), deep learning can be expected to show its generalization power and revolutionize the field of molecular function prediction.
\end{abstract}

\section{Introduction}

Recent developments in experimental techniques for life sciences allow for studying a vast array of properties and characteristics of biologically relevant molecules. We can elucidate structures of proteins and small molecules, their composition (in terms of amino acid sequences and atoms), abundance and localization in the cells, to name just a few such traits. Most of these experimental efforts serve one purpose though -- uncovering the \emph{function} the molecule carries out in the living organism. Both for proteins and small molecules, the function can be described in terms of the effect that the molecule has on its interaction partners. This can in turn be expressed in terms of spatial interactions, such as lock-and-key model of ligand affinity or enzyme specificity (based on spatial complementarity).

While elucidating the structures of biomolecules becomes easier, experimental function annotation remains elusive.  According to UniProtKB, out of over 74~million proteins in the database just 89~thousand have experimentally determined function, 393~thousand have been labelled with a function by a human expert and only \textasciitilde{}12\% (9~million) have been annotated in any form, be it by human or by a computer algorithm.

Analogously, PubChem (one of the largest databases of drug-like molecules and their bioactivity assays) contains over 93 million compounds, but only 1.2~million experimental assays in which one or more of the compounds have been tested against one of \textasciitilde{}10~thousand protein targets or \textasciitilde{}20~thousand gene targets. Bearing in mind that a single compound can act on multiple targets and a single protein can be a target of many compounds, it is evident that this database is far from being comprehensive.

\subsection{Non-structure-based function prediction}

While the function is dictated by structure, it is not always necessary to know the full, atomic structure of the compound to be able to infer the function. For example, protein function prediction methods use numerous sources of information in combination. In addition to the amino acid sequence (primary structure), one can use evolutionary information (homologs of known function), sequence information inferred from genome (genomic context), gene co-expression, proteomic assays (including protein-protein interaction), as well as data from genetic assays and clinical observations, as summarized in an overview of numerous state-of-the-art methods~\citep{CAFA2}.

\subsection{Structure-based function prediction}

\subsubsection{Related work}

There are many possibilities to represent information about molecule structure, and to feed it into a function prediction method.

For quantitative structure-activity relationship (QSAR) modeling, chemical structures are often numerically encoded with hand-made descriptors that describe chemical properties, topology or atomic connectivity, and spatial geometry of the molecule~\citep{sliwoski_computational_2013}. Scalar descriptors include molecular properties, such as molecular weight and the octanol-water partition coefficient (LogP). Topological descriptors encode the connectivity of the molecule, examples of which include substructure-matching schemes and bond distance histograms such as 2D autocorrelation functions~\citep{sliwoski_autocorrelation_2016}.  Geometrical descriptors include radial distribution and 3D autocorrelation functions which calculate histograms of interatomic distances within a molecule, or encoding coefficients of the spherical harmonics which best describe the shape of the molecule~\citep{baumann_distance_2002,wang_spherical_2011}. Topological and geometrical descriptors are often weighted by atomic properties such as partial charge or polarizability to describe the spatial and topological distributions of these properties as well. Most of descriptors either do not require a three-dimensional conformation of the molecule, or else a single low-energy conformation is used for their calculation.  Four-dimensional descriptors have also been described which aim to encode some dynamical properties of molecules, such as multiple conformations~\citep{andrade_4d-qsar_2010}, in addition to the properties described above.

For proteins, on the other hand, approaches to representing the structure are two-fold, either coordinate-based or topology-based. The first ones denote the positions of all the amino acids in Cartesian space, either by coordinates of atoms or of pseudoatoms (larger entities representing a group of atoms). Coordinate-based representations are immediately interpretable, as they comprise sufficient information to easily position all the objects in three-dimensional space. The functional relationships within the protein, though, are better captured by representations taking into account the mutual proximity of the objects (amino acids, atoms\ldots). Such approaches have been widely adopted in the field, ranging from directly enumerating distances between bodies (often within a certain cutoff), through enumerating the bodies that are in spatial proximity (in contact, within a certain distance threshold) according to a certain metric, to purely neighborhood-based measures (such as the ones dictated by Voronoi tesselation)~\citep{voro3d}.  Through these measures it is straightforward to tell which bodies (atoms, amino acids\ldots) interact, but exceedingly difficult to reconstruct the original structure, if the original distances have not been preserved.

The method most similar to ours -- using 3D representations directly as inputs to the neural network -- is AtomNet~\citep{AtomNet}. Its details and differences to our method are described below in Section~\ref{atomnet-differences}.

\newpage
\subsubsection{Motivation for proposed structure-based method}

At a microscopic level, the electromagnetic force governs interactions between molecules of any shape or size and in particular it is responsible for the binding affinities of small molecules to proteins or for enzyme function.  A classical description of molecular structure usually differentiates two major subsets of electromagnetic interactions, namely electrostatic forces (often described by an electrostatic potential) and van der Waals or steric interactions~\citep{israelachvili_intermolecular_2011}. Electrostatic forces are longer-range attractive or repulsive effects which are a result of charge imbalances between atoms, that establish partial positive and negative charges in different regions of the molecular structure.  
Van der Waals interactions are a shorter-range effect which may be either attractive, due to transient dipoles in the electron clouds, or repulsive, due to an overlap of the electron clouds of molecules.  Van der Waals effects are therefore determined by the electron-dense regions around a molecule which effectively determine its shape and play an important role in determining binding interactions.  Variations of electrostatic potential and electron densities are often used to computationally describe molecular interactions at a classical level, such as in molecular dynamics calculations~\citep{salomon-ferrer_overview_2013,brooks_charmm_1983,alper_computer_1989}.  Together these two properties make up a majority of what two interaction partners ``see'' of each other. In other words, electron density (or its estimate) and electrostatic potential are major determining factors of the molecular function. 
This is why we propose using these fields directly as inputs to the function prediction method.

\begin{figure}[!tbp]
\centerline{\includegraphics[width=\columnwidth]{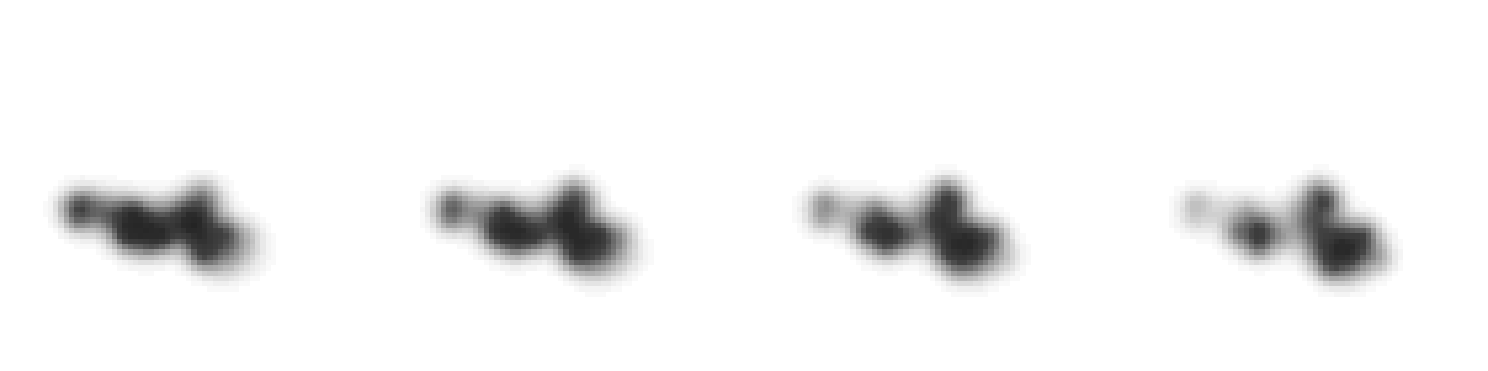}}
\centerline{\includegraphics[width=\columnwidth]{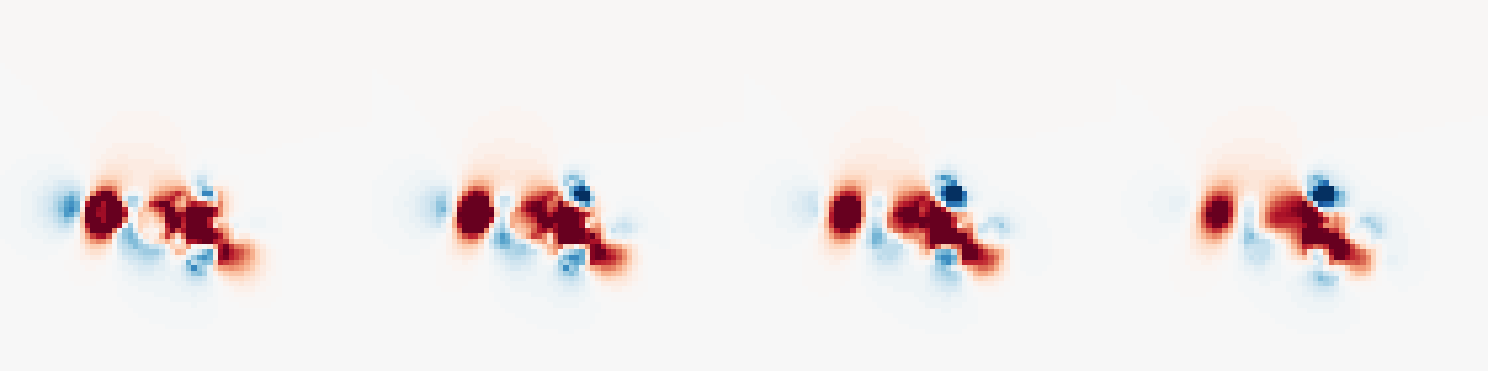}}
\caption{Four of 70 $z$-slices from the two-channel $70 \times 70 \times 70 \times 2$ representation of an active M$_1$ muscarinic receptor agonist (from dataset PubChem SAID 1798): approximated electron density field (top) and electrostatic potential field (bottom; positive potential in blue, negative in red). These physical fields characterize how this molecule can spatially fit to other molecules (binding partners). We thus propose using these fields directly as input to the 3D convolutional network.}
\label{fig:2channels}
\end{figure}

The common theme in related work is that nearly all methods that use known (or predicted) structural information for function prediction extract handcrafted features. These transformations discard part of the information contained in the original data, very likely to the detriment of subsequent analysis. Lessons learned from the success of deep learning in numerous areas of application~\citep{AlexNet,NIPS,UltraDeepContact} consistently indicate that deep learning can deal with the entire known raw information and learn the data transformation that is optimal for the task at hand. The multi-layer (deep) data transformations applied to the raw data are optimized jointly in view of the final goal (such as classification), formulated as the cost function of the neural network. The output error is back-propagated through all neural network layers, i.e.\ optimizes the transformations at all layers. In most cases, such automatically optimized transformations strongly outperform handcrafted ones. We therefore explore the deep learning approach to molecular function prediction.

\subsubsection{Differences between AtomNet and proposed method}
\label{atomnet-differences}

To our knowledge, the only similar method is AtomNet~\citep{AtomNet}. It uses various 3D grid representations of small molecules for bioactivity prediction. The main differences between AtomNet and our approach are the following:
\begin{itemize}
\item AtomNet requires a 3D co-complex of the molecule in question with its binding target, whereas our method infers the function of the molecule from the shape of the molecule alone.
\item The version of AtomNet that uses the enumeration of atom types in a 3D grid only implicitly provides the information about the possibilities for physical interactions with other molecules, whereas our representation of the molecule via its approximate electron density and electrostatic potential 3D grids is a rawer, more direct representation of how other molecules ``perceive'' the molecule in question, and in what ways they can physically interact. Besides, the enumeration of atom types in a discrete grid without anti-aliasing introduces imprecisions, partially discarding information about the exact relative positions of the atoms, whereas the usage of anti-aliasing for atom enumeration or the voxel-wise estimation of electron density and electrostatic potential are unaffected by discretization-based imprecisions.
\item The version of AtomNet that uses handcrafted chemical descriptors such as SPLIF~\citep{SPLIF}, SIFt~\citep{SIFt}, or APIF~\citep{APIF} brings along the aforementioned disadvantages of handcrafted features, whereas we provide the entire physical information required to extract relevant chemical and physical properties.
\item Besides small molecules, we also present \emph{protein} function prediction using 3D deep learning, demonstrating the robustness of our approach. 
\item The network architecture of our approach differs from the one of AtomNet in the following ways: (a) it contains max-pooling layers, thus encouraging the learning of invariances (cf.\ below); (b) our network has more convolutional layers. The overall architecture is based on the work of \cite{VGG}.
\end{itemize}

\section{Methods}

\subsection{Input representation}

Electron density of a molecule cannot be easily computed based on the coordinates alone, but can be approximated based on the positions of atoms and their van der Waals radii. Approximate electron density and electrostatic potential are calculated on a Cartesian grid. For small-molecule experiments, the field of view is $\SI{35}{\angstrom} \times \SI{35}{\angstrom} \times \SI{35}{\angstrom}$ at a resolution of $\SI{0.5}{\angstrom}$, i.e.\ $71 \times 71 \times 71$ voxels. An example is shown in Fig.~\ref{fig:2channels}. For protein experiments, the field of view is $\SI{127}{\angstrom} \times \SI{127}{\angstrom} \times \SI{127}{\angstrom}$ at a resolution of $\SI{2}{\angstrom}$, i.e.\ $64 \times 64 \times 64$ voxels, for low-resolution experiments, and a resolution of $\SI{1}{\angstrom}$, i.e.\ $128 \times 128 \times 128$ voxels, for high-resolution experiments.

The approximate electron density used herein is emphatically \emph{not} the true, noisy one measured by X-ray crystallography or electron microscopy. The experimental data can be used directly, but this remains to be the subject of future research. Instead, we use an estimate of the idealized electron density obtained from the atom coordinates. The electron density is estimated using a Gaussian kernel around the atom center with the van der Waals radius as its bandwidth. Using electron density from physical experiments directly is in principle also possible.

The electrostatic potential of small molecules is estimated by computing the partial charges of the atoms using the Gasteiger-Marsili  PEOE (Partial Equalization of Orbital Electronegativities) algorithm~\citep{gm78}. 

Protein experiments were performed using approximated electron density only, without electrostatic potential, for several reasons. Firstly, the computation of partial charges for proteins requires employing one of several non-trivial algorithms, all of which are fraught with substantial degree of error. The comparison of the effect of these methods on the results of 3D deep learning is subject of future work. Secondly, individual amino acid types have a paramount effect on the partial charges, and amino acid types can be inferred from the approximated electron density alone. Thus, due to the limited vocabulary of structural motifs in proteins, ``electrostatic motifs'' can be learned from ``electron density motifs''.

\begin{figure}[!tbp]
\centerline{\includegraphics[width=\textwidth]{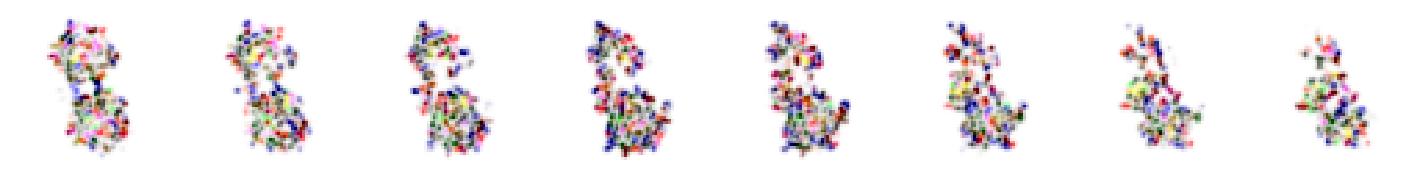}}
\caption{Eight $z$-slices of approximate protein electron density field in a 21-channel representation. The backbone and each residue type have individual channels (shown as 21 colors) for the electron density of respective atoms. This novel protein-specific representation helps the neural network to distinguish the amino acid types directly.}
\label{fig:21channels}
\end{figure}

To additionally simplify the recognition of structural motifs and to make use of the limited vocabulary of amino acid residues, we propose an alternative \textbf{multi-channel input representation}. Instead of computing the approximated electron density of the entire protein in a Cartesian $64 \times 64 \times 64$ voxel grid (or $128 \times 128 \times 128$ for high-resolution experiments), we separate the atoms by the 20 amino acid residue types they belong to and calculate the approximated electron density (Gaussian kernel) 3D maps for each residue type separately, yielding 20 channels, i.e.\ a $64 \times 64 \times 64 \times 20$ array. An additional channel is used for the electron density of backbone atoms, the 21 channels thus summing up to the overall approximated electron density. These 21 channels are shown in Fig.~\ref{fig:21channels}. Furthermore, one additional channel is used for hydrogen atoms due to their role in hydrogen bonds, influencing molecular function, and a complementary channel for heavy-atom electron density, the two latter channels also summing up to the overall approximated electron density. Finally, another channel holds the approximated electron density for all atoms, providing three-fold redundancy, but also a disentangled information representation amenable to learning relevant deep feature extractors. The overall input size is thus $64 \times 64 \times 64 \times 24$. The three spatial dimensions are used for 3D convolutional layers, and the fourth dimension represents the channels, i.e.\ the voxel-wise features.

We encourage rotation-invariance by training, i.e.\ by using \emph{data augmentation}, in this case \textbf{random rotations and translations} of the molecule when creating the Cartesian grid of the physical fields. This trains the neural network to produce similar output for a certain molecule regardless of its position and orientation in space. Furthermore, data augmentation prevents overfitting and facilitates generalization, since unimportant (``overfittable'') features such as a specific orientation of the molecule (and associated local ``voxel value motifs'') are never repeated during training, whereas structural invariants relevant for predicting molecular function are maintained.

\begin{figure}[!tbp]
\centerline{\includegraphics[width=\textwidth]{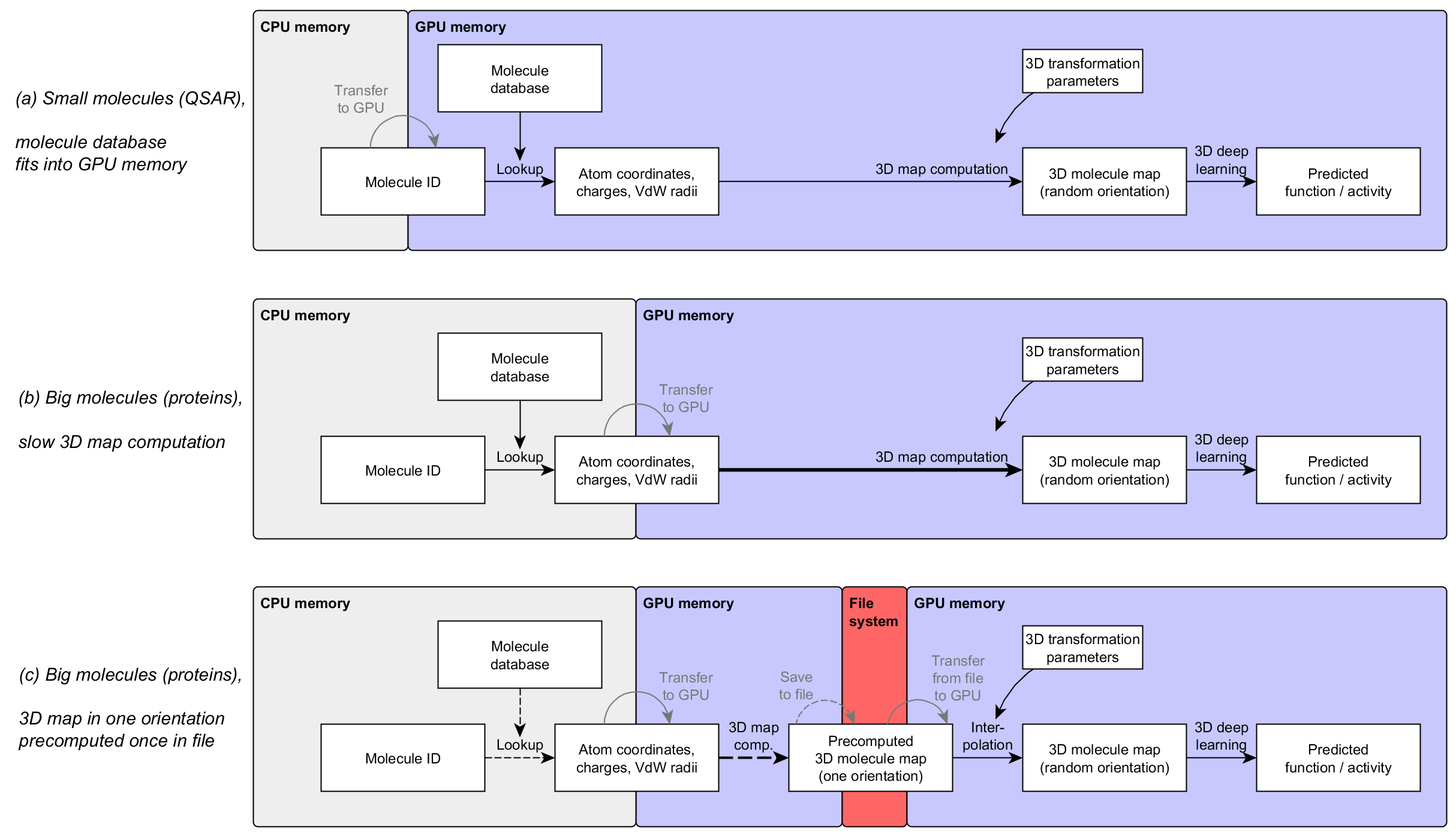}}
\caption{We propose three versions of the pipeline, addressing different needs. The all-GPU pipeline~(a) is appropriate if the molecule database fits into GPU memory, for example in the case of small molecules. The large-database pipeline~(b) can be applied to the study of proteins; its bottleneck is the slow computation of 3D maps. To circumvent this bottleneck, the \emph{fast} large-database pipeline~(c) precomputes the 3D maps only once.}
\label{fig:pipeline}
\end{figure}

\subsection{Input computation pipeline}
\label{sec:pipeline}

Where possible, we perform computations on the graphical processing unit (GPU) due to speed concerns -- not only the deep learning training, but also the computation of the inputs (3D maps). In our implementation with Lasagne~\citep{Lasagne} and Theano~\citep{Theano} software libraries, data generation and data augmentation are seamlessly integrated into the processing as ``layers'' of the neural network.

In cases of large molecules and/or large numbers of molecules, the database may not fit into GPU memory. We therefore propose different pipelines with the molecule database in the GPU memory (Fig.~\ref{fig:pipeline}a) or CPU memory (Fig.~\ref{fig:pipeline}b--c). Moreover, the computation of 3D maps is a bottleneck in case of large molecules. Thus, we also propose a pipeline where the 3D maps are pre-computed in one orientation, stored in a file, and rotated for purposes of data augmentation during training (Fig.~\ref{fig:pipeline}c), resulting in slight interpolation artifacts, but retaining the important physical information. 

We use the all-GPU pipeline (Fig.~\ref{fig:pipeline}a) for QSAR and the fast large-molecule pipeline (Fig.~\ref{fig:pipeline}c) for protein function prediction.

\subsection{Neural network architecture}

Protein function is dictated by the shape of the active site (as represented in the physical 3D maps), as well as the folds (evolutionary families) and relative positions of the domains of the protein. The overall structure of active sites and domains can be inferred from local structural motifs and their higher-level global composition.

The method should therefore have the following properties:

\begin{enumerate}
\item Translation-covariance of low-level feature extraction
\item Locality of low-level feature extraction
\item Hierarchical feature extraction from localized and simple to larger and more abstract
\item Rotation-invariance
\end{enumerate}

The first three points are ensured by employing convolutional neural networks. The fourth point is taken care of by random rotations during training.

The four-dimensional (3D $\times$ channels) feature maps $\mathbf{a}^{(l)}$ in a 3D convolutional layer $l$ are computed according to the operation
\begin{equation}
a^{(l)}_{xyzc} = \sigma\left(b^{(l)}_{c} + \sum_{\hat{x}\hat{y}\hat{z}k} w^{(l)}_{\hat{x}\hat{y}\hat{z}kc} a^{(l-1)}_{x+\hat{x},y+\hat{y},z+\hat{z},k}\right),
\end{equation}
where $\mathbf{a}^{(l-1)}$ are the activations of the previous layer, $\mathbf{b}^{(l)}$ the biases and $\mathbf{W}^{(l)}$ the weights trained by error backpropagation such that the objective function is optimized, and $\sigma$ is the nonlinearity, in our case the leaky rectified linear unit \citep{leakyReLU} defined as $\sigma(z)=\mathrm{max}\{0.01x, x\}$.

\begin{figure}[!tbp]
\centerline{\includegraphics[width=0.8\columnwidth]{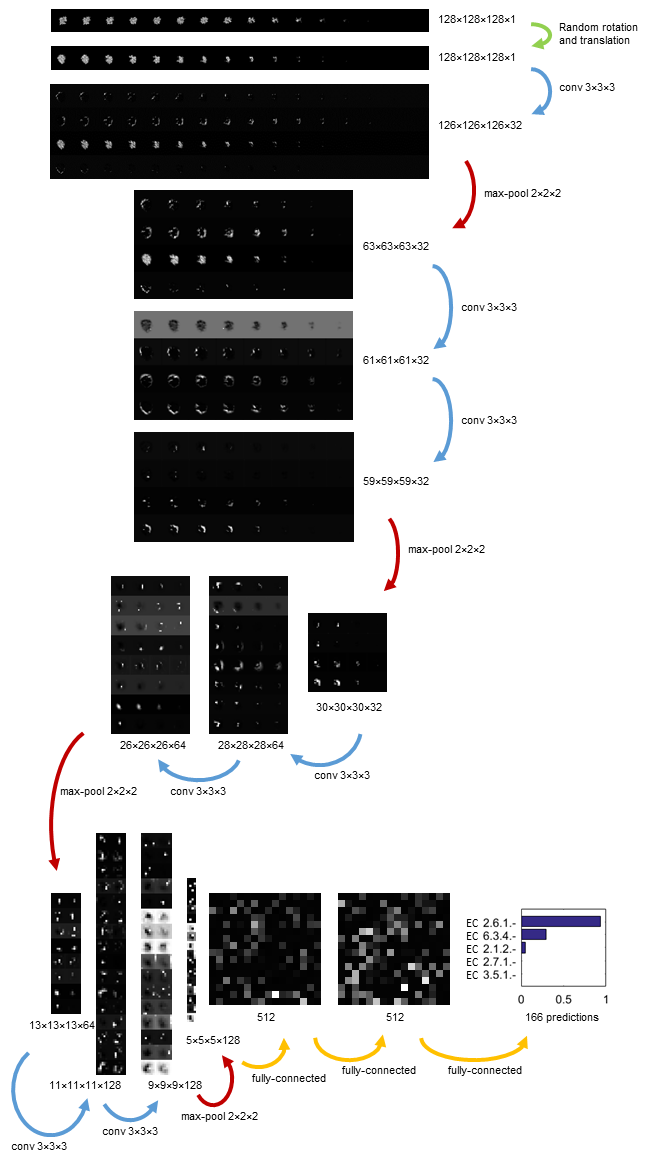}}
\caption{Schematic of the neural network architecture used in most experiments, with visualized activations (feature maps) after training. Only a fraction of slices and channels is shown (with consistency across layers). The outputs do not have to sum up to 1 because one sample can belong to several classes (and therefore sigmoid rather than softmax was used as the nonlinearity in the output layer).}
\label{fig:network}
\end{figure}

The neural network architecture is closely based on a design practice popularized by~\cite{VGG}, proposing the usage of very small convolutional filters, achieving certain receptive window sizes through increased depth of the network (allowing more elaborate data transformations) rather than large filters. Pooling layers increase the receptive window size; they encourage (but do not enforce) the learning of invariance to slight distance changes, slight translations and slight rotations; and they contribute to a higher level of feature abstraction, model regularity (generalizability) and computational tractability.
We use 3D convolutional filters of size $3 \times 3 \times 3$, interspersed with 3D pooling layers of size $2 \times 2 \times 2$, analogously to 2D operations in the work of~\cite{VGG}.
The network architecture and activation maps are shown in Fig.~\ref{fig:network}.

The network has an output unit for each predicted class. The output nonlinearity is the sigmoid
\begin{equation}
\sigma(z) = \frac{1}{1+e^{-z}}
\end{equation}
(rather than the softmax function common in classification), so that molecules with several functions can be modeled in a straightforward manner by setting \emph{several} output targets to 1 for the same sample (where 0 or 1 represents class membership). The corresponding loss (objective) function for $C$ classes is the binary cross-entropy between predictions $\mathbf{y}$ and respective targets $\mathbf{t}$, summed for all classes:
\begin{equation}
L = \sum_{c=1}^{C} -t_c \log y_c - (1-t_c)\log (1-y_c).
\end{equation}

\begin{figure}[!tb]
    \centering
    \begin{subfigure}[b]{0.49\textwidth}
        \includegraphics[width=\textwidth]{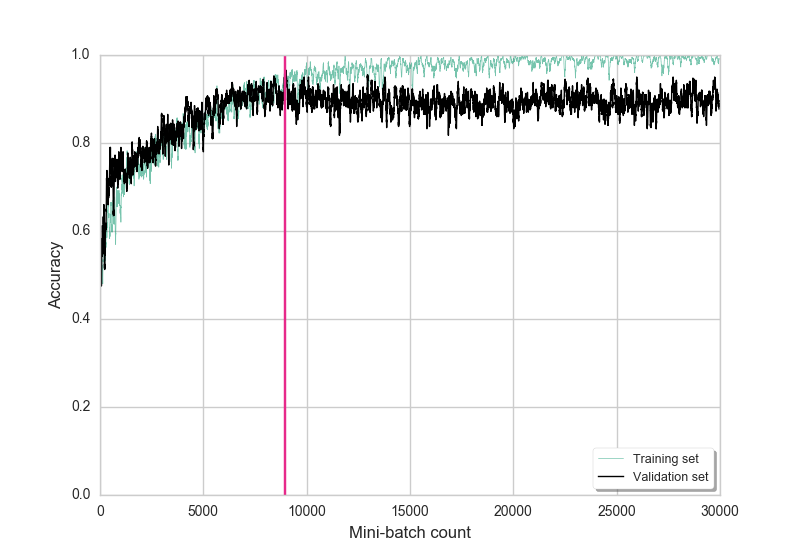}
        \caption{Validation accuracy history}
        \label{fig:accuracy}
    \end{subfigure}
    \begin{subfigure}[b]{0.49\textwidth}
        \includegraphics[width=\textwidth]{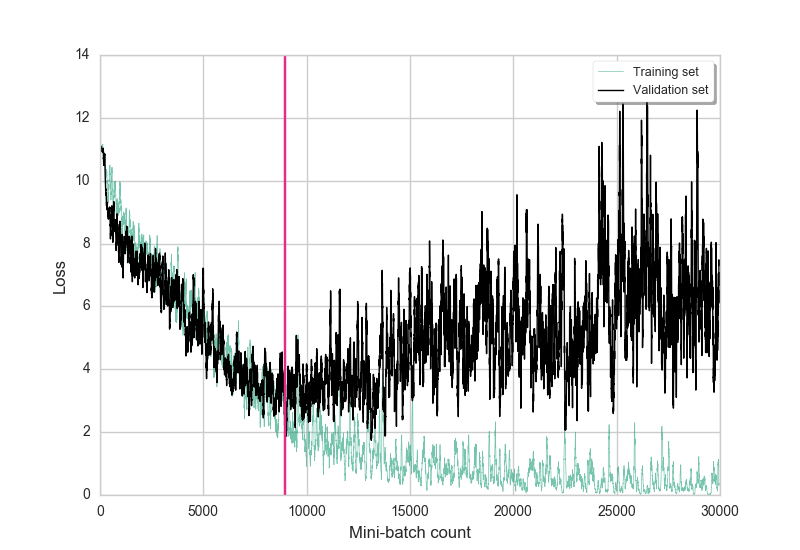}
        \caption{Validation loss history}
        \label{fig:loss}
    \end{subfigure}
    \caption{Accuracy history~(a) for protein function prediction: EC 3.4.21.- (serine proteases) vs.\ EC 3.4.24.- (cysteine proteases). Early stopping (magenta line) is performed when the all-time average validation set accuracy stops improving.
    This early stopping criterion is meaningful also in terms of validation loss~(b), avoiding overfitting and variance.}
    \label{fig:accuracy_loss}
\end{figure}

Training is performed using the Adam algorithm \citep{Adam} with a learning rate of $10^{-4}$, mini-batch size 2, 4 or 8 for different resolutions and GPUs. Early stopping (Fig.~\ref{fig:accuracy}) is performed when the average validation accuracy stops improving; the average is taken since the beginning of the training.

We use the VoxNet implementation~\citep{Maturana2015,VoxNet} of 3D convolutional and 3D pooling layers. Generation and augmentation of 3D molecule maps are directly integrated with the Lasagne/Theano-based neural network training (cf.\ Section~\ref{sec:pipeline} and Fig.~\ref{fig:pipeline}).

\subsection{Data}

For protein function prediction, we use the Enzyme Structures Database\footnote{{https://www.ebi.ac.uk/thornton-srv/databases/enzymes/}}, which lists PDB structures for enzymes classified by the  EC (Enzyme Commission) number hierarchy, designating enzyme specificity and mode of action.  

In the first experiment, we perform classification on two classes of enzymes acting on a peptide bond (peptidases): serine proteases (EC~3.4.21.-) and cysteine proteases (EC~3.4.24.-). Proteins in both classes perform the same task of cleaving a peptide bond, but differ in terms of catalytic mechanism. Both classes contain proteins that do not necessarily share evolutionary history (i.e.\ are not necessarily homologous) and therefore do not always share the same structure (fold). The characteristic that proteins within either of the classes share is the same functional mechanism, which may have emerged in the course of convergent evolution.

In another larger-scale experiment, we perform classification of all 166 third-level EC classes against each other, i.e.\ EC~1.1.1.-  vs.\ EC~1.1.2.- vs.\ \dots\ vs. EC~6.6.1.-. We experiment with two different methods of splitting samples into training and test sets. By \emph{naive split}, we refer to randomly assigning samples (PDB entries) to the training or test set. A more challenging task is function prediction with a \emph{strict split} at the fourth EC level, meaning that if e.g.\ a sample in EC~3.4.21.1 (chymotrypsin) gets randomly assigned to the test set, then all other entries in this fourth-level class also get assigned to the test set and none of them to the training set. Thus, it is tested whether the neural network can correctly predict chymotrypsins (EC~3.4.21.1) to belong to the class of serine proteases (EC~3.4.21.-), based solely on information inferred from \emph{other} subclasses of EC~3.4.21.-, such as subtilisin, thrombin or trypsin, but no samples from chymotrypsin class.

In each training/test split, we select 25\% of the samples for the test set (consistently across experiments of the same splitting method). Additional 25\% are picked randomly as a validation set to perform early stopping during training in order to prevent overfitting (cf.\ Figs.~\ref{fig:accuracy}--\ref{fig:loss}). The remaining 50\% of the data are used for training.

For the small-molecule QSAR task, we use a dataset of M$_1$ muscarinic receptor agonists and inactive componds (PubChem SAID 1798) that were annotated with their activity in a respective assay. We attempt to classify the molecules into the active/inactive categories, reserving 20\% of the data for testing, and training on the other 80\%.

\begin{figure}[!b]
    \centering
    \begin{subfigure}[b]{0.49\textwidth}
        \includegraphics[width=\textwidth]{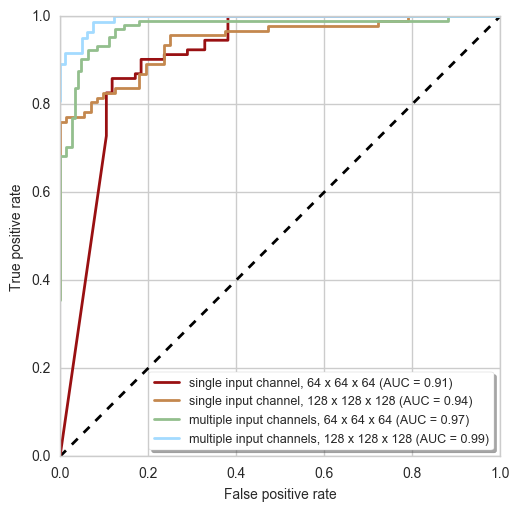}
        \caption{Naive split}
        \label{fig:roc:naive}
    \end{subfigure}
    \begin{subfigure}[b]{0.49\textwidth}
        \includegraphics[width=\textwidth]{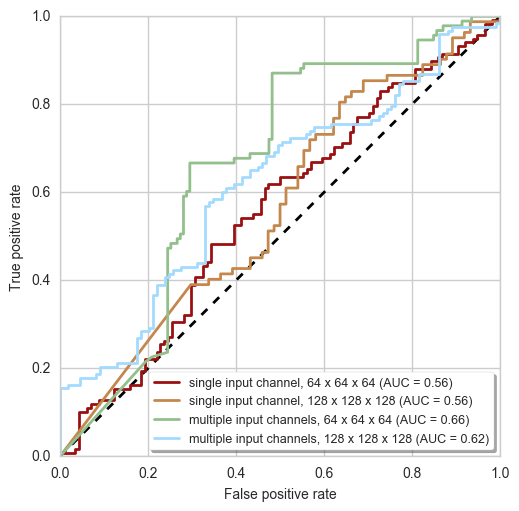}
        \caption{Strict split}
        \label{fig:roc:strict}
    \end{subfigure}
    \caption{ROC for protein function prediction: EC~3.4.21.- (vs.\ EC~3.4.24.-) with naive split~(a) and strict split~(b). The multi-channel input representation outperforms the respective single-channel settings. The proposed 3D input representation provides meaningful information about the molecules under the naive split and even under the challenging strict split.}
    \label{fig:roc}
\end{figure}

\section{Results}

The receiver operating characteristic (ROC) for discriminating between serine and cysteine proteases, i.e.\ classifying EC~3.4.21.- against EC~3.4.24.- with the naive training/test split is shown in Fig.~\ref{fig:roc:naive}. For low-resolution experiments, the area under the ROC curve (AUC) of 0.91 with single-channel inputs and 0.97 with multi-channel inputs indicates a high quality of prediction and suggests that \emph{high accuracy protein function prediction from 3D maps is feasible}. Moreover, doubling the resolution in each dimension, i.e.\ using $128 \times 128 \times 128$ voxels in lieu of $64 \times 64 \times 64$ voxels, further increases the AUC to 0.94 for single-channel and 0.99 for multi-channel experiments. Thus, both of the high-resolution settings outperform the corresponding low-resolution settings; and both of the multi-channel settings outperform the respective single-channel setting. This indicates that the high-resolution data representation and the multi-channel data representation (and both together) contain valuable information for this task.
However, the naive split entails that both test and training set most probably contain proteins from the same evolutionary family, thus making the learning task substantially easier.

ROC for the same classes with a strict training/test split is shown in Fig.~\ref{fig:roc:strict}. The AUC=0.56 for single-channel inputs is much lower than for the naive split, confirming that the strict split -- inferring protein function only based on \emph{other} protein families sharing same function -- is a considerably harder problem. However, the AUC=0.66 for multi-channel inputs also shows that the \emph{multi-channel representation of protein structure is beneficial} for facilitating the extraction of information about protein function.

With the strict split, increasing the resolution of the data does not strongly influence the AUC, and in case of the multi-channel inputs leads to decrease in predictive power. It is however important to note that increasing the resolution notably improves the expected number of true positives before encountering a false positive (left part of ROC curve; for both single- and multi-channel methods with naive split, as well as for single-channel for strict split). This signifies that the network given improved resolution learns to recognize crucial structural features of individual classes. The subsequent, mid-range drop in accuracy can, in our opinion, be attributed to insufficient amount of data in the training set, which prevents the network from generalizing properly. This is also the reason for sub-par performance of the method in ``high resolution, multi-channel'' regime.

It is however evident from Figures~\ref{fig:roc:naive} and~\ref{fig:roc:strict} that introduction of additional channels denoting the amino acid types substantially improves the expected prediction accuracy. In our opinion, this allows the network to easier attain the knowledge on structural and functional properties of amino acids, which it would have to learn in training. By introducing additional channels, we alleviate this need. It is evident that analogous effect could have been achieved by pretraining the networks with a larger data corpus, which we will demonstrate in subsequent work. Therefore, we postulate that in the limit of infinite (or at least sufficiently abundant) data, performance of ``single-channel'' and ``multi-channel'' approaches would converge.

\begin{figure}[!tpb]
\centerline{\includegraphics[width=\columnwidth]{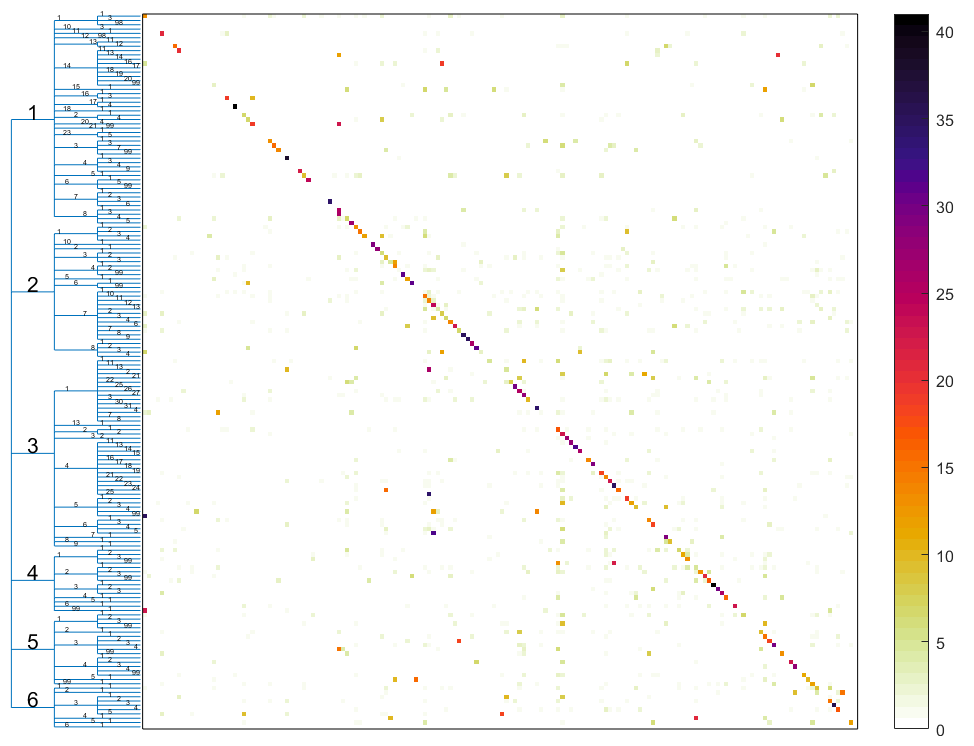}}
\caption{Confusion matrix for classification of the 166 third-level EC classes. The pronounced diagonal means correct prediction for the majority of test samples.}
\label{fig:confmat}
\end{figure}

The confusion matrix for the classification of the 166 third-level EC classes with naive train/test split is shown in Fig.~\ref{fig:confmat}. The diagonal (correct predictions) is highly pronounced, indicating that our deep learning approach stably distinguishes the correct class from the 165 other possible classes of proteins in many cases, which is a challenging task requiring the representation of numerous function-specific structural features within the neural network.
Each class had a slightly different AUC in the test data; the AUC of the individual classes averaged across all classes was $0.87 \pm 0.13$ on the test data.

\begin{figure}[tpb]
    \centering
    \begin{subfigure}[b]{0.53\textwidth}
        \includegraphics[width=\textwidth]{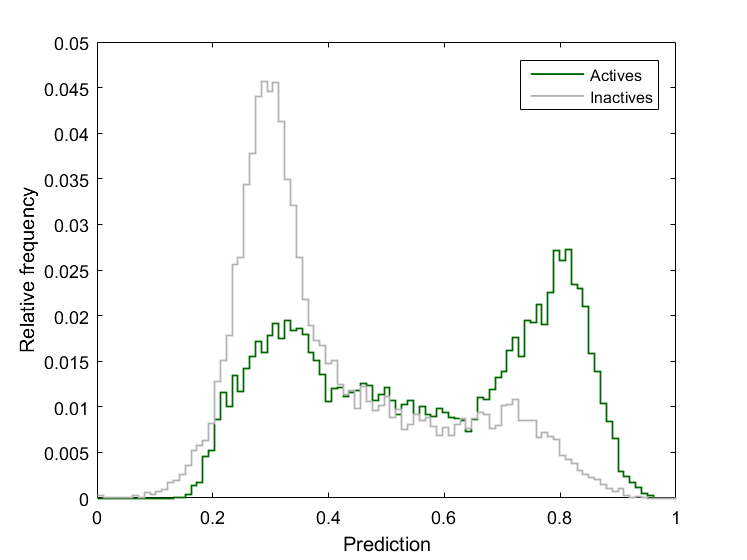}
        \caption{Normalized histogram of predictions (neural network outputs) for active and inactive small molecules}
        \label{fig:qsar:histogram}
    \end{subfigure}
    \quad
    \begin{subfigure}[b]{0.42\textwidth}
        \includegraphics[width=\textwidth]{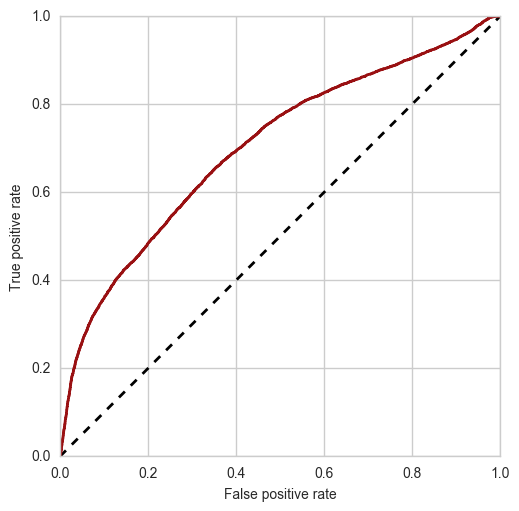}
        \caption{ROC (AUC=0.70)}
        \label{fig:qsar:roc}
    \end{subfigure}
    \caption{Results for small-molecule QSAR. Predictions~(a) demonstrate some separability of active and inactive small molecules from the test set. The ROC~(b) has an AUC=0.70 approximately equal to that of state-of-the-art descriptor-based methods applied to this dataset (PubChem SAID 1798).}
    \label{fig:qsar}
\end{figure}

The ROC curve for biological activity classification of M$_1$ muscarinic receptor antagonists is shown in Figure~\ref{fig:qsar:roc}.  Models trained on this dataset achieved an AUC=0.70.  This value indicates that the models were capable of differentiating molecules which exhibited biological activity from those that did not at a rate substantially higher than random chance.  The AUC value found here is \emph{approximately equal to that of state-of-the-art methods based on hand-crafted descriptors} applied to the same dataset~\citep{mendenhall2016}.  A further demonstration of the performance of the model can be seen in Figure~\ref{fig:qsar:histogram}, which highlights a substantial gap between mode scores for the two classes of compounds.

These results are both interesting and encouraging given that state-of-the-art models utilize descriptors and network architectures that are specifically optimized for biological activity prediction.  The performance of the models reported here indicates that the chosen approximations for describing molecular structure are already capable of encoding information that is equally valuable to what is found in the hand-crafted descriptors.  Further refinement of both molecular structure description and network parameters are likely to boost the performance above what is seen here.  Additionally, these models illustrate that deep learning can be a powerful tool even in domains where the data sets are much smaller and more heavily unbalanced than those seen in more traditional applications of deep learning.

As noted in \citep{mendenhall2016}, the goal of biological activity prediction is often to prioritize a large set of compounds in order to select a small subset for physical testing, thereby making those compounds with the highest scores the most important from a practical standpoint.  These models were trained with a loss function designed to optimize the overall AUC of the ROC curves (i.e.\ from false positive rate (FPR) values of 0 to 1) which effectively aims to best separate the two classes from each other as a whole.  However, this metric does not consider the performance of the highest scoring samples, and as seen in Figure~\ref{fig:qsar:histogram}, the highest-scoring active compound is separated from the highest-scoring inactive compound by a small fraction of the score range.  Given these promising early results, approaching this problem with a loss function designed to optimize compound scores at low FPR values (e.g.\ the logAUC metric described in \citep{mendenhall2016}) would provide a straightforward way to improve the performance of these models in a manner beneficial for practical QSAR application.

\section{Discussion and Conclusions}

In this work we have demonstrated the utility of 3D convolutional neural networks for discriminating between molecules in terms of their function in living organisms. They allow for accurate classification of proteins, without relying on domain (expert) knowledge or evolutionary data. Additionally, use of the 3D convolutional neural networks for classifying small molecules permits for achieving comparable degree of accuracy as state-of-the-art QSAR methods, while obviating the need for handcrafted descriptors. 

The data on protein classification presented above was based purely on enzymes, but it is highly plausible that these results generalize to the entire protein space. The success in discriminating between vastly different enzyme classes strongly indicates that this approach allows for high-confidence discrimination between different functional classes (transporters, structural proteins, immunoglobulins\ldots). Discriminating within a class (e.g.\ dopamine receptor from a glucagon receptor) should be possible as well, as demonstrated by serine vs.\ cysteine protease experiment.

The classification experiments presented above are to be treated as proof-of-concept. While we demonstrated the applicability of convolutional neural networks for these purposes, we do acknowledge that the predictive power of these methods can only increase when supplemented with additional features, used by the other methods in the respective fields. For purposes of protein function prediction, one could easily improve expected prediction accuracy by extending the feature set by such ones as presence of common sequence motifs, structural classification (annotated or predicted) and homology-derived information. The small-molecule classification can be augmented by experimentally determined features (e.g.\ logP, polarizability\ldots) and ones derived from the structure (e.g.\ constitutional descriptors, fragment counts\ldots). While the latter can be learned from the data itself, it may prove beneficial to provide them explicitly.

Notably, this work does not consider the fact that the vast majority of molecules in living organisms is conformationally flexible. It is possible to generate multiple conformers of small molecules, for the QSAR use case, but it is not intuitively obvious what effect will it have on the training of the method. By using generated conformations as actives one would inadvertently introduce false positives in the training set (i.e.\ conformations in which the ligand does not bind would be labeled as positives), but on the other hand it would allow the network to recognize also potentially active ligands, even if they were in an unsuitable conformation. 
For protein function prediction, conformational flexibility plays a less major role, but distinction between apo (without bound ligand) and holo (with ligand bound) structures in the training process could potentially play a role for the expected predictive power.

While the protein function prediction part of this work relied on experimental structural data, 
these methods can also be applied to \emph{predicted} protein structures. It could be advisable to limit the prediction to the active site only, thus allowing for much faster training and predictions, and enabling meta-prediction using multiple variant active sites. 

The other potential use case is to use experimental electron density directly, without the need for fitting atoms within. Recent developments in the area of direct imaging, especially electron microscopy, make such methods particularly relevant. As in its current form our method does not depend on any sequence-related data, it is immediately applicable to such problems. This could enable high-confidence function annotation of proteins recovered from environmental samples.

These are just a few of potential application domains for the methods we propose. By avoiding human-derived, handcrafted descriptors they allow to capture the features of the studied molecules that are truly important for functional considerations. On contrary to these descriptors, they will only increase in accuracy with the growing amount of data. On contrary to methods based on structural comparison, 
the methods we proposed do not require superposition and are translation and rotation invariant. We postulate therefore that deep learning methods of inferring functional information from raw spatial 3D data will increase in importance, with the growing amounts of spatial biological information and increased resolutions of direct imaging methods.

\section*{Funding}
This work was supported by the European Research Council [Consolidator Grant ``3DReloaded'' to V.G. and D.C.]; and by the Deutsche Telekom Foundation.

\bibliographystyle{plainnat}

\end{document}